# A flexible film thermocouple temperature sensor

Yulong Bao, Bin Xu, Huang Wang, Dandan Yuan, Xiaoxiao Yan, Haoxin Shu(B), and Gang Tang(B)

Jiangxi Province Key Laboratory of Precision Drive and Control, Nanchang Institute of Technology, Nanchang 330099, China

`2916579466@qq.com, tanggangnit@163.com`

## Abstract

This article introduces a thin-film thermocouple temperature sensor with symmetrical electrode structure. It uses PI film as a flexible substrate. Cu film and CuNi film made by MEMS manufacturing process as positive and negative electrodes. The device itself has the advantages of miniature, bendable and fast response speed. To reduce the film resistance value. Conducting metal film thickness and sputtering substrate temperature optimization experiments. The critical dimensions of Cu/CuNi film are 650nm and 400nm. The best sputtering substrate temperature for Cu/CuNi films is 100°C and 150°C. Testing the adhesion of thin film thermocouples using the peel-off method. The test result is 9.4N. Finally, the film thermocouple temperature sensor is subjected to a temperature static calibration experiment. The result shows that the actual potential difference error is within ±1°C. It belongs to the second-class standard in the formulation of thermocouple standards in my country. Through curve fitting, the corresponding relationship between temperature and potential difference is more accurate.

**Keywords**: thin film thermocouple, sensor, MEMS process, static calibration

## 1. Introduction

With the rapid development of science and technology, various types of film materials are widely used in production and life [1-4]. The thin film thermocouple temperature sensor is a new type of sensor born with the development of thin film technology [5-7]. Compared with the traditional bulk thermocouple, due to its small device and the thickness of the thermal junction at the micro-nano level, it has the advantages of fast response and small heat capacity [8-9].

In this work, we propose a thin-film thermocouple temperature sensor based on PI flexible substrate to achieve flexibility and rapid temperature measurement.

Design appropriate size and structure, use MEMS technology to prepare finished products and conduct experimental tests to verify the performance of the finished products.

## 2. Working principle and thermoelectric performance analysis

2.1 working principle

Thermocouple temperature sensor is a widely used temperature measurement device, and its temperature measurement principle is thermoelectric effect [10]. The thermoelectric effect refers to the thermoelectric phenomenon that occurs due to the mutual contact of different types of metals, that is, two different metals form a closed loop. When there is a temperature difference between the two connectors, current will be generated in the loop.

2.2 Thermoelectric performance

The physical quantity that characterizes a thermocouple is the thermoelectric potential rate, which is essentially the relative thermoelectric properties of the two different materials that make up the thermocouple. The thermoelectric properties of a single material are called absolute thermoelectric potential rate or absolute Seebeck coefficient [11]. defined as:

$$S = \int_0^T \frac{\mu}{T} dT \tag{1}$$

in the formula, $\mu$ ——Thomson coefficient of the material
$T$ ——absolute temperature

This definition of absolute thermoelectric potential rate is derived from the Kelvin relational formula, which is collated as:

$$S_{AB} = \int_0^T \frac{\mu_A}{T} dT - \int_0^T \frac{\mu_B}{T} dT \tag{2}$$

Putting equation (1) into the above equation, we can see:

$$S_{AB} = S_A - S_B \tag{3}$$

It can be seen from formula (3) that the Seebeck coefficient $S_{AB}$ of a thermocouple temperature sensor is equal to the algebraic sum of the absolute thermoelectric potential rates of the two hot electrode materials that make up the thermocouple.

## 3. Device design and manufacturing

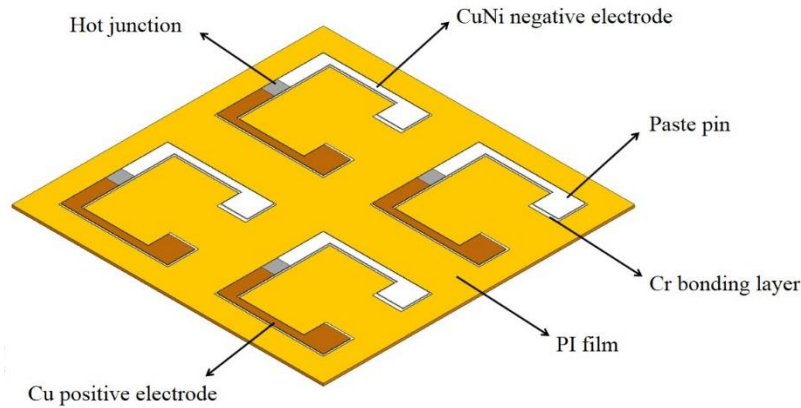

Figure 2. Schematic diagram of the structure of thin film thermocouple temperature sensor

The designed thin film thermocouple temperature sensor is shown in Figure 1. The structure includes PI film substrate, bonding layer Cr, positive electrode Cu, negative electrode CuNi, and a thermal junction composed of two electrodes in contact. The overall size of the device is 24mm×24mm, and there are 4 thin-film thermocouple temperature sensors of the same size, each of which is relatively independent. The Cu and CuNi film structure is designed to be a symmetrical structure, and the size of the thermal junction is 1mm×1mm. The size of the compensation wire pin is 2mm×2mm, which is convenient for bonding the lead. The bonding layer Cr is located between the PI film substrate and the electrode layer, the width at the hot junction is increased by 0.5mm, and the connection pin is increased by 0.5mm. The purpose is to prevent deviations during preparation, so that the Cu and CuNi electrodes are not completely sputtered on the bonding layer Cr, causing the thermocouple electrode to fall off.

The process flow chart of the thin-film thermocouple temperature sensor prepared by MEMS technology is shown in Figure 2. Paste the PI film on a clean silicon wafer and perform the cleaning and drying process, as shown in Figure 2(a); Carry out the leveling process, and evenly spin-coat a layer of positive photoresist AZ4620 with a thickness of 1 μm on the PI surface, as shown in Figure 2(b); Use a lithography machine to perform the exposure process, and the pattern designed on the lithography, as shown in Figure 2(c); Use high vacuum three-target magnetron coating system for sputtering process to make the device sputter a layer of Cr metal film as a whole, as shown in Figure 2(d); Use Lift-off process to remove the remaining photoresist, leaving a complete bonding layer Cr pattern, as shown in Figure 2 (e). The positive electrode Cu and the negative electrode CuNi are manufactured according to the MEMS manufacturing process in the above steps. Finally, paste the Cu compensation wire to the Cu electrode pin, and paste the CuNi compensation wire to the CuNi electrode pin.

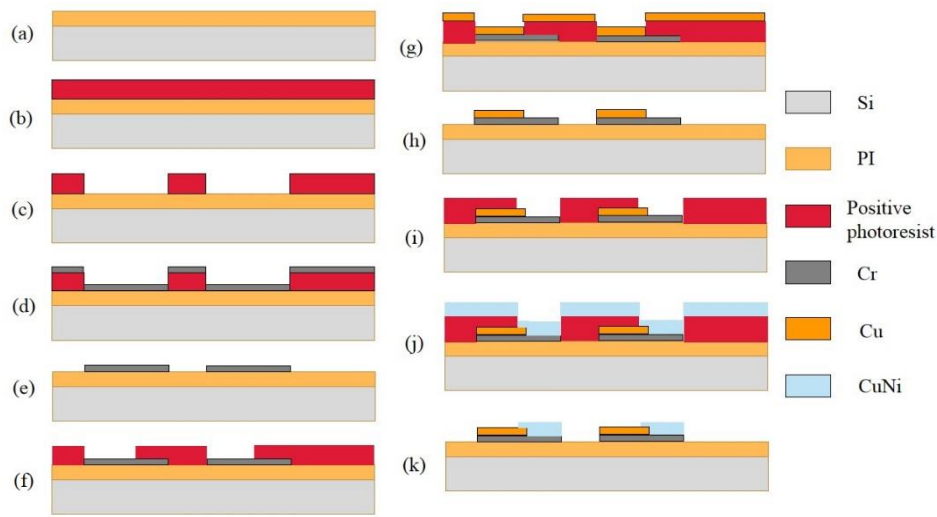

Figure 2. Process flow chart of thin film thermocouple temperature sensor

## 4. Experimental testing and analysis

4.1 Experimental test platform

The temperature static calibration test platform of the thermocouple temperature sensor is shown in Figure 3. The instruments include: a constant temperature heating platform and a KEITHLEY electrometer. The constant temperature heating table can control the temperature stably, the temperature control range is between normal temperature and 350 ℃, the control accuracy is accurate and the performance is stable, and the error is within ±1℃. The KEITHLEY electrometer has voltage measurement parameters, which can be used for potential difference testing to obtain the relationship between temperature and potential difference.

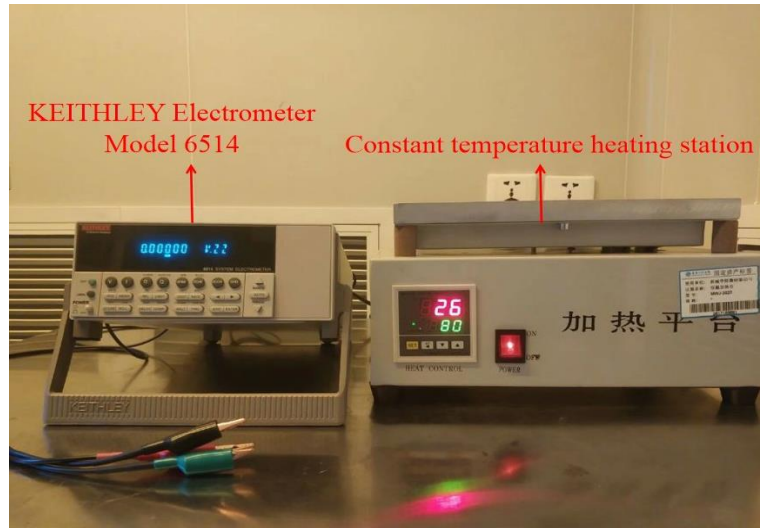

Figure 3. Temperature static calibration experiment platform

4.2 Metal film thickness optimization experiment

When a metal conductor has a temperature gradient, it will cause the number of electrons near the Fermi energy at both ends of the conductor to change with energy, with more ultra-high-energy electrons on the high-temperature side and more ultra-low-energy electrons on the low-temperature side. When conducting electrons diffuse in a conductor, their diffusion rate is related to energy: the higher the energy of electrons, the less scattering they will receive during diffusion, and the greater the diffusion rate. A net diffused electron flow will be formed in the conductor, so that conduction electrons will accumulate at one end of the conductor, generating an electromotive force. The thermoelectromotive force generated by the diffusion of electrons due to the existence of the conductor temperature difference is called "diffusion thermoelectromotive force". The derivative of temperature is called absolute thermoelectric potential rate [12].

When the thickness of the metal thin film or the alloy thin film is small to a certain extent, the scattering probability of the conductive electrons by the boundary increases, which affects its diffusion thermoelectromotive force rate. As a result, the thermoelectric properties of the thin film are different from solid materials. This phenomenon is called the "film size effect".

The thickness of the sputtered metal film is different, which will affect the film quality and resistivity of the thermocouple temperature sensor. The "film size effect" shows that the resistivity of the metal film will increase sharply when the thickness is less than a certain thickness. The rate will be basically stable, and this thickness is called the critical dimension.

The experimental results of film thickness optimization are shown in Figure 4. It can be seen from Figure 4(a) that when the thickness of the Cu film is between 350nm and 650nm, the resistance value decreases rapidly from 12.8Ω to 3.5Ω. After 650nm, the resistance value tends to stabilize, so it can be judged that the critical thickness of the Cu electrode is 650nm. It can be seen from Figure 4(b) that when the film thickness is between 250nm and 400nm, the resistance value rapidly decreases from 180Ω to 53.2Ω. After 400nm, the resistance value tends to stabilize, so the critical thickness of CuNi electrode is judged to be 400nm.

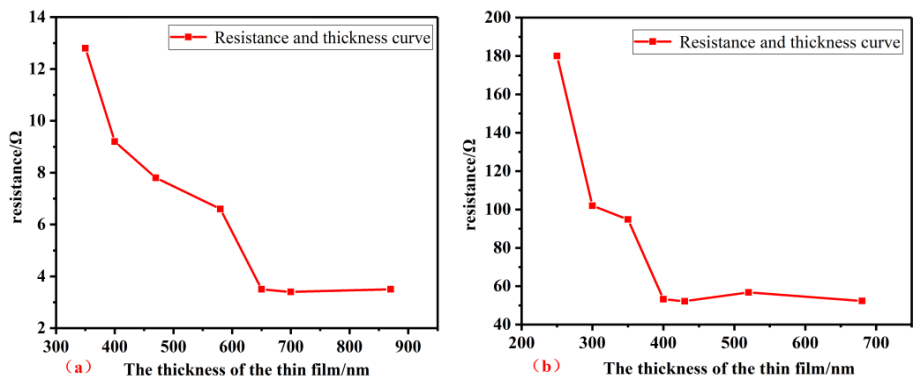

Figure 4. Cu/CuNi film resistance and thickness relationship curve

4.3 Temperature optimization experiment of metal film sputtering substrate

Properly increasing the temperature of the substrate during sputtering can not only reduce the film stress and enhance the bonding force of the film substrate, but also reduce the film resistivity and increase the thermoelectromotive force.

The experimental results of the temperature optimization of the metal thin film sputtering substrate are shown in Figure 5. It can be seen from Figure 5(a) that the resistance value drops rapidly from 3.5Ω to 1.1Ω when the temperature is 30℃ ~ 100℃, and the resistance value between 100℃ ~ 200℃ is relatively stable. It is inferred that the optimal Cu film sputtering The substrate temperature is 100°C. It can be seen from Figure 5(b) that the resistance value drops rapidly from 53.2Ω to

32.4Ω at a temperature of 30°C to 150°C. Although the resistance value is also falling at 150°C to 200°C, the overall resistance value tends to be stable, so Inferred that the optimal CuNi film sputtering substrate temperature is 150℃.

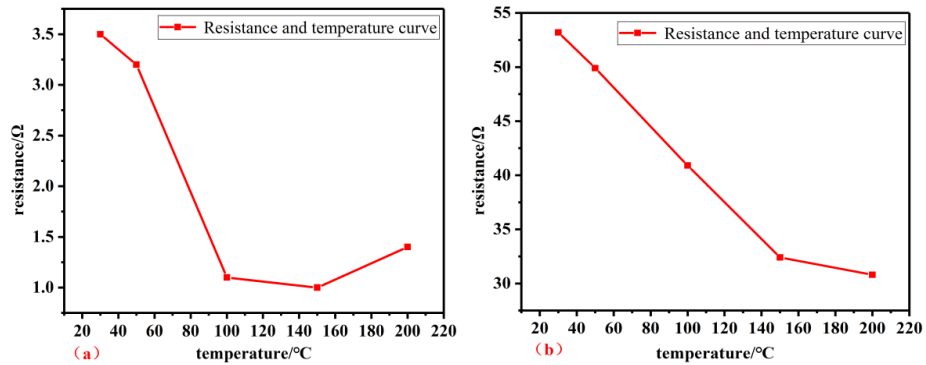

Figure 5. Cu/CuNi film resistance and temperature curve

4.4 Adhesion test experiment

The adhesion between the film and the substrate directly affects the performance of the film, and films with poor adhesion cannot be applied to flexible devices. Internal stress is generated during the film preparation process, and internal stress is also generated between the flexible substrate and the metal film. Excessive internal stress may cause the film to curl or crack, or the film will fall off when the flexible substrate is torn. Therefore, the adhesion of the film determines the possibility and reliability of the application of flexible devices.

Adhesion is one of the important indicators to describe the performance of the film. The adhesion test methods include two types: one is the adhesive method, including the pull method, the peel method, and the pull method; the other is the direct method, including the scratch method, the friction method, Centrifugation etc. This paper uses the peeling method to test the adhesion between the film and the flexible substrate. A length of PI tape is torn out, one end is fixed with a clip, and the other end is pasted on the Cu/CuNi electrode of the thermocouple temperature sensor to cover the entire electrode layer. Pull the tension sensor of the push-pull dynamometer upward, observe the change of the reading and the peeling of the film by the tape, and stop if there is a small amount of film peeling off. After testing, it is concluded that when the applied pressure is about 9.4N, the metal film begins to fall off slightly, which shows that the adhesion force is 9.4N.

### 4.5 Thermocouple static calibration experiment

According to the temperature static calibration test platform built in Figure 3, the thermocouple temperature sensor is statically calibrated. First, paste the thermocouple temperature sensor on the silicon chip, put it on the constant temperature heating table and fix it with PI tape. The test cable of the KEITHLEY electrometer clamps the compensation lead of the positive and negative electrodes respectively. Because there may be an error between the temperature displayed by the constant temperature heating station and the surface temperature of the thermocouple temperature sensor, it is necessary to calibrate the actual temperature. After testing, the error between the temperature displayed on the constant temperature heating station and the actual temperature is about 5°C.

The calibration experiment was formally started. A thermometer was used to measure the temperature of the end of the compensating wire, that is, the cold end of the wire to be 22°C. The reference to the standard thermocouple index table shows that the corresponding potential difference is 0.87mV. The constant temperature heating stage starts heating at 22°C, the set temperature interval is 1°C, and it heats from 22°C to 80°C and records the potential difference per 1°C. According to the law of thermocouple intermediate temperature, it can be known that the actual electric potential difference must be added to the cold junction electric potential difference of 0.87mV. The comparison curve between the actual thermocouple electric potential difference and the standard electric potential difference is shown in Figure 6. The analysis shows that the actual electric potential difference error is ±1℃, which is in the standard of thermocouples in my country It is a level II standard under development.

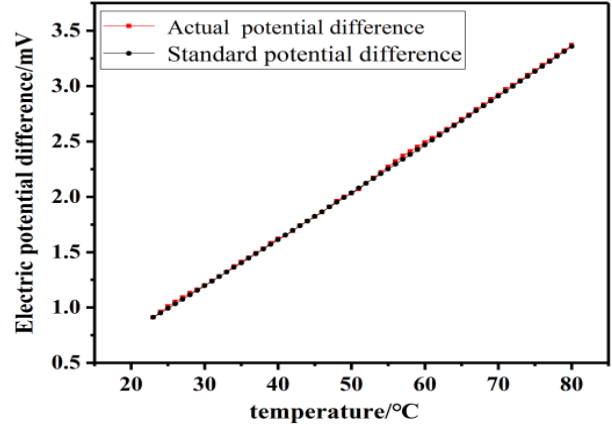

Figure 6. Comparison curve between actual potential difference and standard potential difference

Because the temperature interval set in the thermocouple static calibration experiment is 1 ℃, the potential difference value cannot further express more accurate temperature changes, so this article fits the actual thermocouple potential difference and the temperature change curve to make it at 23 ℃ ~ 80 ℃ Any potential difference value within the range corresponds to a temperature value. The fitting curve is shown in Figure 7. The fitting relationship is a linear equation, and the fitting formula is:

$$y = 0.04309x - 0.09827 \tag{4}$$
$$R^2 = 0.99963$$

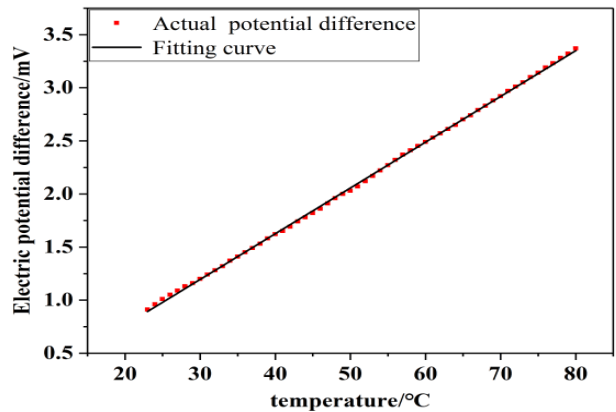

Figure 7. Fitting curve of relationship between actual potential difference and temperature change

## 5. Conclusion

In this paper, a T-type thermocouple temperature sensor with a symmetrical electrode structure is designed, using PI film material as a flexible substrate, which is flexible and bendable for use. Manufactured by MEMS manufacturing process, the device is more miniature and precise. In order to reduce the resistance value of the metal film, an optimization experiment for the thickness of the metal film and the temperature of the sputtering substrate was set to obtain the critical size and the best sputtering substrate temperature. At the same time, the adhesion test experiment was added, and the adhesion of the film thermocouple temperature sensor was about 9.4 N. Finally, through temperature static calibration experimental research and analysis, the results show that the actual potential difference error is ±1 ℃, which belongs to the second-level standard in the formulation of thermocouple standards in my country. In order to be more precise in the experimental test results, curve fitting was performed on the relationship between the actual thermocouple potential difference and the temperature change.